\documentclass[conference]{IEEEtran}
\usepackage{graphicx}
\usepackage{subfigure}
\usepackage{amsmath}
\usepackage{amssymb}
\usepackage{amsthm}
\usepackage{cite}

\begin{document}

\title{Talk More Listen Less: Energy-Efficient Neighbor Discovery in Wireless Sensor Networks}
\author{
\IEEEauthorblockN{Ying Qiu, Shining Li, Xiangsen Xu and Zhigang Li}
\IEEEauthorblockA{School of Computer Science, Northwestern Polytechnical University, Xi'an, China\\
Email: \{qiuying, xuxiangsen\}@mail.nwpu.edu.cn, \{lishining, lizhigang\}@nwpu.edu.cn}
}
%

\maketitle

\begin{abstract}
Neighbor discovery is a fundamental service for initialization and managing network dynamics in wireless sensor networks and mobile sensing applications.
In this paper, we present a novel design principle named Talk More Listen Less (TMLL) to reduce idle-listening in neighbor discovery protocols by learning the fact that more beacons lead to fewer wakeups.
We propose an extended neighbor discovery model for analyzing wakeup schedules in which beacons are not necessarily placed in the wakeup slots.
Furthermore, we are the first to consider channel occupancy rate in discovery protocols by introducing a new metric to trade off among duty-cycle, latency and channel occupancy rate.
Guided by the TMLL principle, we have designed Nihao, a family of energy-efficient asynchronous neighbor discovery protocols for symmetric and asymmetric cases.
We compared Nihao with existing state of the art protocols via analysis and real-world testbed experiments.
The result shows that Nihao significantly outperforms the others both in theory and practice.
\end{abstract}


\IEEEpeerreviewmaketitle

\section{Introduction}
Wireless networks have greatly changed our daily life in recent years,
using various mobile devices such as smartphones, tablets and smartwatches.
Innovative mobile applications utilize location proximity to provide interesting services,
such as proximity-based social networks and online taxi services (Uber~\cite{uber}, Kuaidi~\cite{kuaidi}).
The first step to connect with another device is to discover each other in the neighborhood.
However, wireless devices are generally battery-powered and suffer from the fact that 
the wireless radio communication and GPS localization are energy-intensive operations.

Energy-efficient neighbor discovery protocol is required for power-constrained mobile devices,
which is extensively researched in the wireless sensor network literature.
Neighbor discovery protocols focus on balancing two contradictive aspects: high energy-efficiency and low discovery latency.
Since the radio operation dominates the energy consumption of the sensor node~\cite{dutta07procrastination},
energy-efficiency is determined by \emph{duty-cycle}, \emph{i.e.} the fraction of time that the radio is on.

Numerous neighbor discovery protocols for wireless sensor networks and mobile networks have been proposed in recent years.
Probability-based approaches, such as Birthday~\cite{mcglynn2001birthday} protocol, randomly select the transmit, receive or sleep state with different probabilities.
In spite of excellent average discovery latency, it exhibits a long tail due to its probabilistic nature. 
To eliminate the long tail with worst-case latency bounded, deterministic discovery protocols are proposed subsequently.
The representative ones are Quorum~\cite{tseng2003power}, Disco~\cite{dutta2008practical}, U-Connect~\cite{kandhalu2010u},
SearchLight~\cite{bakht2012mobicom}, BlindDate~\cite{wang2015blinddate}, Hello~\cite{sun2014hello}, Hedis and Todis~\cite{chen2015infocom},
which have continuously pushed the boundary of neighbor discovery protocol research.

Enlightened by SearchLight and BlindDate, we propose a family of neighbor discovery protocols named Nihao that reduce idle-listening by transmitting more beacons.
We make the following contributions:
\begin{enumerate}
    \item By observing the fact that the 2-beacon approach in Disco leads to redundancy and idle-listening can be reduced by sending more beacons, we propose a principle called Talk More Listen Less (TMLL) for designing better discovery protocols.
    \item We present an extended neighbor discovery model for analyzing discovery protocols that beacons are not placed in the wakeup slots.
    \item We are the first to consider channel occupancy rate in neighbor discovery by introducing the DC-L-COR product metric to formalize and compare existing discovery protocols.
    \item By introducing the notion of balance factor, we design a family of discovery protocols named Nihao that work both in symmetric and asymmetric cases, which are better than the state of the art discovery protocols in theory.
    \item We evaluate Nihao on real-world testbeds. The result shows Nihao significantly outperforms the state of the art protocols both in symmetric and asymmetric cases.
\end{enumerate}

The rest of the paper is organized as follows.
Section~\ref{motivation} explains the motivation of our work and proposes the TMLL principle.
Section~\ref{model} introduces an extended neighbor discovery model for analyzing the wakeup schedules that beacons are not placed in the wakeup slots.
We present the design of a family of Nihao discovery protocols in Section~\ref{design}.
Section~\ref{evaluation} presents an evaluation of Nihao in real-world testbeds.
We discuss the related works in Section~\ref{relatedworks} and finally conclude the paper in Section~\ref{conclusion}.

\section{Motivation}\label{motivation}
This section explains the motivation of our work. We will prove redundant beacon exists in Disco's 2-beacon approach for bidirectional discovery.
Then, inspired by the fact that SearchLight and BlindDate utilize the redundant beacons to reduce wakeups,
we propose a more general principle to guide neighbor discovery protocol design.

\subsection{Redundant Beacon}\label{redundantbeaconsection}
Slot model is commonly used in analyzing the performance of neighbor discovery protocols.
In slot model, time is divided into slots with identical length. If slots are perfectly aligned, a discovery is guaranteed when both nodes are active in the same slot.
However, slots are rarely aligned in practice since nodes run asynchronously and precise global time synchronization is hard to achieve.
To ensure bidirectional discovery when slots are not aligned, Disco~\cite{dutta2008practical} sends a beacon at both the beginning and end of an active slot.
This method is widely adopted by existing discovery protocols, such as SearchLight~\cite{bakht2012mobicom}, BlindDate~\cite{wang2015blinddate} and Hello~\cite{sun2014hello}.

\begin{figure}[t]
    \centering
    \includegraphics[scale=0.6]{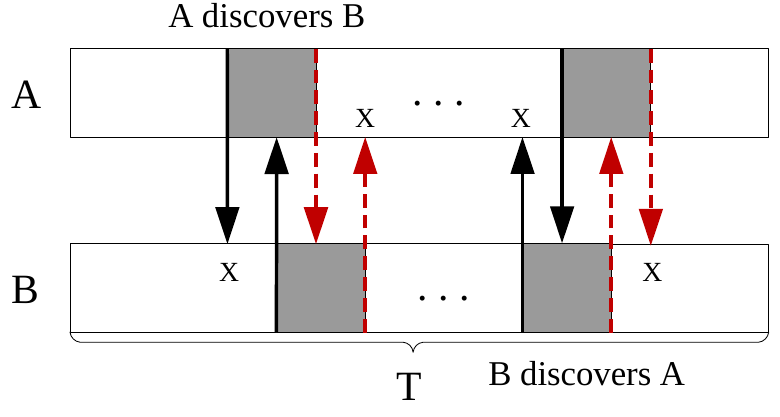}
    \caption{Redundant beacons in Disco. Two discoveries with different directions are guaranteed in a schedule cycle regardless of the offset.
    The beacons with dashed arrows are redundant for bidirectional discovery.}\label{redundantbeacon}
\end{figure}

However, we argue that the 2-beacon approach is not energy-efficient enough because one beacon is sufficient for bidirectional discovery.
Eliminating the beacon at the end of an active slot will not affect bidirectional discovery.
The reason is that Disco only notices a partial overlap of two slots but ignores the other part of the overlap.
Figure~\ref{redundantbeacon} illustrates why the 2-beacon approach proposed by Disco is redundant.
Removing the beacon at the end of an active slot (with dashed arrows in figure) will not affect bidirectional discovery and the worst-case latency bound,
but will slightly decrease the discovery rate.

\begin{figure}[t]
    \centering
    \includegraphics[scale=0.6]{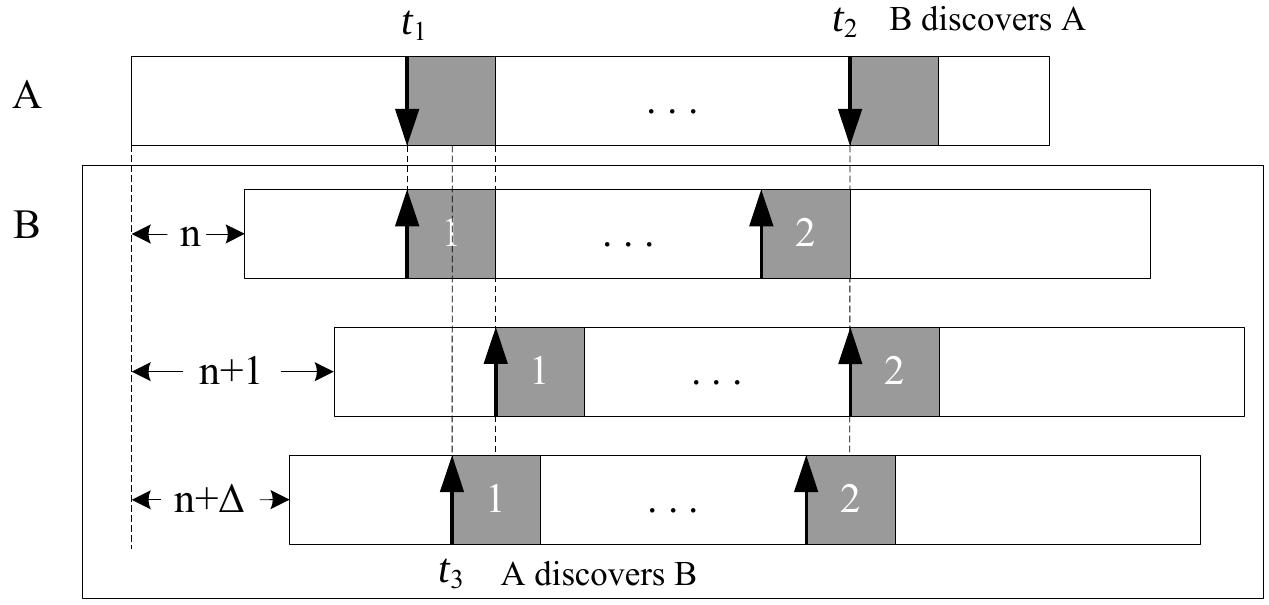}
    \caption{Proof: one beacon is sufficient for bidirectional discovery.}\label{redundantbeaconproof}
\end{figure}

Now we prove \emph{one beacon in an active slot is enough for bidirectional discovery} in deterministic discovery protocols without striped probing.
\begin{proof}
Deterministic discovery protocol guarantees at least one overlap active slot in a schedule cycle.
When slots are aligned and the offset between node A and B's wakeup schedule is $n$, there must be at least one overlap at slot $t_1$.
Similarly, when the offset is $n+1$, the overlap slot is $t_2$.
Now, we consider the case that slots are not aligned, suppose the offset is $n+\Delta$, where $0 < \Delta < 1 $.
Since A, B still overlap with part of each other, A can discover B on $t_3=t_1+\Delta$, while B can discover A on $t_2$. Therefore, bidirectional discovery is guaranteed.
\end{proof}
Figure \ref{redundantbeaconproof} illustrates the proof in detail. The experiment result in section~\ref{redundantbeaconexpsection} also meets our proof.

\subsection{Talk More Listen Less Principle}
Existing quorum-based neighbors discovery protocols use a column of \emph{anchor} slots overlap with a row of \emph{probe} slots to guarantee discovery in a matrix schedule.
SearchLight first notices that the 2-beacon approach leads to two discovery opportunities when slot boundaries are not aligned.
With only probing every even slot, SearchLight eliminates half of the probe slots in a wakeup schedule.
Furthermore, BlindDate eliminates more probe slots by adding two extra beacons to the beginning of previous slot and the end of next slot, 
achieving a better worst-case bound than SearchLight.
These approaches have provided important insights that \emph{adding more beacons can reduce probe slots},
especially considering the fact that one beacon is sufficient for bidirectional discovery proved above.
Meanwhile, BlindDate have inspired us that \emph{beacons are not necessarily placed in the wakeup slot}.
We believe these are not just workarounds for non-alignment of the slots, but a certain pattern in common exists.

\begin{figure}[t]
    \centering

    \subfigure[Quorum: 1 beacon, full probes] {
        \includegraphics[scale=0.45]{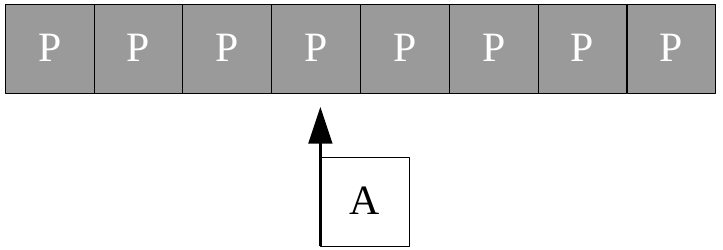}\label{morebeaconlessprobe1}
    }\hspace{1em}
    \subfigure[SearchLight: 2 beacons, 1/2 probes] {
        \includegraphics[scale=0.45]{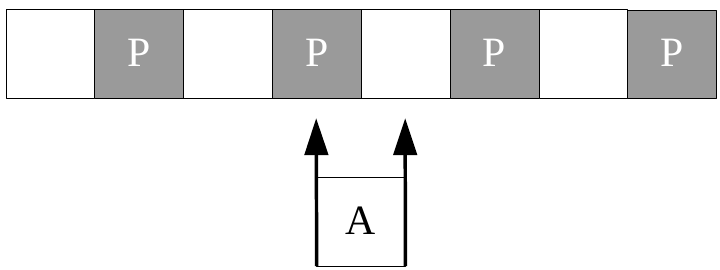}\label{morebeaconlessprobe2}
    }
    \subfigure[BlindDate: 4 beacons, 1/4 probes] {
        \includegraphics[scale=0.45]{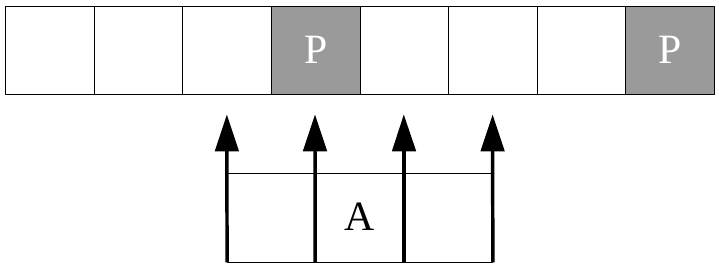}\label{morebeaconlessprobe3}
    }\hspace{1em}
    \subfigure[Full beacons, 1 probe] {
        \includegraphics[scale=0.45]{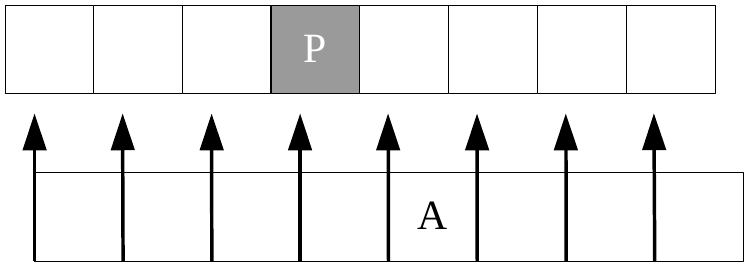}\label{morebeaconlessprobe4}
    }
    \caption{More beacons lead to fewer probes.}\label{morebeaconlessprobe}
\end{figure}

As Fig.~\ref{morebeaconlessprobe} illustrates, when beacons increase, fewer probes are necessary for discovery.
Since node in active slot should keep the radio on for the duration of the whole slot, it wastes more energy than just transmitting a single beacon.
Utilizing the fact that a short beacon is more energy-efficient than an active slot can significantly reduce idle-listening.
We summarize the \emph{Talk More Listen Less} (TMLL) principle which indicates to reduce wakeup slots by sending more beacons:

\begin{itemize}
    \item since the beacon is shorter than the wakeup slot, adding more beacons will reduce the number of probe slots;
    \item beacons are not necessarily placed in the wakeup slot.
\end{itemize}

To better explain TMLL, we first propose the Talk-Listen model in next section which is well-suited for discovery schedules that wakeup slots and beacons are separated.

\section{Model}\label{model}
In this section, we present the Talk-Listen model, which extends the traditional model used by most existing discovery protocols.

\begin{figure}[t]
    \centering
    \includegraphics[scale=0.5]{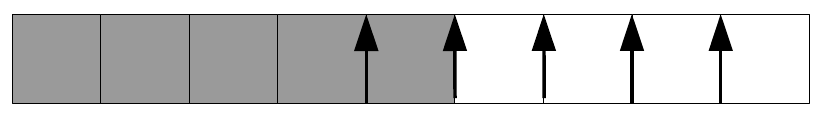}
    \caption{An example of discovery schedule that cannot fit into Listen-Listen model.
             Beacons and wakeup slots take up half of a cycle respectively, ensuring at least two overlaps.}\label{examplellcanntexplain}
\end{figure}

Traditional neighbor discovery model (Listen-Listen model) only cares about the overlap of two wakeup slots, so it fails to model discovery schedules that beacons are separated from wakeup slots.
For example, Fig.~\ref{examplellcanntexplain} is a legal schedule that ensures neighbors will discover each other in one cycle.
Listen-Listen model cannot represent the beacons that are not in the wakeup slot, which is expressible in our Talk-Listen model.

\subsection{Discovery Schedule}
In Talk-Listen model, wakeup slots and beacons are independently scheduled.
We use similar notations in U-Connect~\cite{kandhalu2010u} to define the discovery schedules of Talk-Listen model.

The discovery schedule of a node $m$ in Talk-Listen model is defined as two binary functions:
$\psi_L (m,t)$ and $\psi_B (m,t)$,
representing the schedule of wakeup slots and beacons at time $t$ respectively:
\begin{eqnarray}
    \psi_L (m,t) & = & \left\{ \begin{array}{ll}
            1, & \textrm{listen for a slot} \nonumber \\
            0, & \textrm{sleep}\\
        \end{array} \right.
\\
    \psi_B (m,t) & = &\left\{ \begin{array}{ll}
            1, & \textrm{send a beacon and back to sleep} \nonumber \\
            0, & \textrm{sleep}\\
        \end{array} \right.
\end{eqnarray}

Neighbor discovery can be defined with $\psi_L (m,t)$ and $\psi_B (m,t)$.
If node $m_1$, $m_2$ can directly communicate with each other, 
a unidirectional neighbor discovery that $m_1$ discover $m_2$ is defined to happen when $\exists t | \psi_L(m_1, t)=\psi_B(m_2, t)=1$.
A bidirectional neighbor discovery that $m_1$ and $m_2$ discover each other is defined \emph{iff}:
\begin{eqnarray}\label{bidiscoverydef}
    \exists t_1,t_2 \quad \psi_L(m_1, t_1) & = & \psi_B(m_2, t_1)=1  \quad \textrm{and} \nonumber \\
                      \psi_B(m_1, t_2) & = & \psi_L(m_2, t_2)=1
\end{eqnarray}

By defining neighbor discovery, a periodic discovery schedules with cycle $T$ must satisfy:
\begin{eqnarray}
    \psi_L (m,t) & = & \psi_L (m,T+t) \nonumber \\
    \psi_B (m,t) & = & \psi_B (m,T+t) \quad \textrm{for all $0 \leq t < T$}
\end{eqnarray}

\subsection{Evaluation Metrics}

Since applications that involve neighbor discovery primarily concern about energy-efficiency and discovery rate.
Key metrics considered by existing discovery protocols are duty-cycle (DC) and worst-case discovery latency (L).

With our Talk-Listen model, the duty-cycle $DC$ of a given periodic discovery schedule with period $T$ is:
\begin{displaymath}
    DC = \frac{1}{T} ( \sum_{t=0}^{T-1} \psi_L(m,t) + \alpha ((\sum_{t=0}^{T-1} \cdot \psi_B(m, t)) - N_c) )
\end{displaymath}
$N_c$ represents the number of common active slots that satisfy $\psi_L(m,t)=1$ and $\psi_B(m,t)=1$.
Subtracting $N_c$ is to avoid double counting when wakeup and beacon occur in the same slot.
$\alpha$ is the proportion between the time for transmitting a beacon and the duration of a wake-up slot.
None of existing discovery protocols has considered $\alpha$, since each beacon is placed inside the wakeup slot.
However, $\alpha$ cannot be ignored when beacons are separated with wakeup slot in Talk-Listen model.
Although the beacon is a short packet that can be broadcasted in less than 1ms with an IEEE 802.15.4-compatible radio,
it will dominate the duty-cycle especially when the TMLL principle is aggressively adopted.

For the convenience of comparing the performance of different protocols, U-Connect proposes the power-latency product metric $\Lambda$:
\begin{displaymath}
    \Lambda = DC \cdot L
\end{displaymath}

As power-latency product cannot directly reflect the overall performance of discovery protocols,
U-Connect introduces the theoretically optimal discovery schedule as the reference point.
Since the optimal schedule is generated by a combinatorial algorithm~\cite{zheng2003asynchronous} under the Listen-Listen model,
we refer to it as LL-Optimal protocol in the following text.
The power-latency product for LL-Optimal $\Lambda_o$ is:
\begin{displaymath}
    \Lambda_o = \sqrt{L-\frac{3}{4}} + \frac{1}{2} \approx \sqrt{L}
\end{displaymath}
Since $\sqrt{L-\frac{3}{4}} + \frac{1}{2} - \sqrt{L} < \frac{1}{2}$ always holds when $L \geq 1$,
we consider $\Lambda_o$ is $\sqrt{L}$ and define $\lambda$ as:
\begin{displaymath}
    \lambda = \frac{\Lambda_{\phantom{1}}}{\Lambda_o} = \frac{\Lambda}{\sqrt{L}}
\end{displaymath}

Besides redefining the above metrics that already exist in Listen-Listen model,
we will introduce a new metric called \emph{channel occupancy rate}, which is described in section~\ref{channeloccupancy}.

\section{Design}\label{design}
In this section, we present the design of the Nihao neighbor discovery protocol under the Talk-Listen model,
which has better performance both in theory and practice.
We begin with a simplified version of Nihao that minimizes the power-latency product.
Then, we take the channel occupancy rate (COR) into consideration and propose the revised Nihao whose COR is adjustable.

\subsection{Simplified Nihao}
\begin{figure}[t]
    \centering
    \includegraphics[scale=0.5]{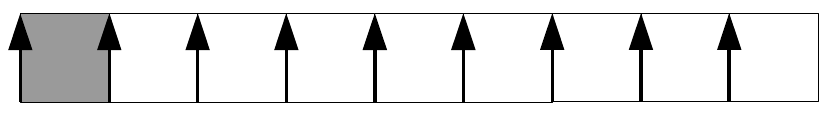}
    \caption{The discovery schedule of S-Nihao. Beacons are transmitted in each slot with only one slot for listening.}\label{simpnihao}
\end{figure}

This section presents the design of the simplified Nihao neighbor discovery protocol (S-Nihao).
Guided by the TMLL principle, we send beacons as more as possible in S-Nihao to reduce probe slots.
As Fig.~\ref{simpnihao} shows, S-Nihao sends a beacon at the beginning of each slot, and only wakes up in the first slot of each schedule cycle.

The discovery schedule of S-Nihao can be represented as:
\begin{eqnarray}
    \psi_L (m,t) & = & \left\{ \begin{array}{ll}
            1, & \textrm{if $[t]_T = 0$} \nonumber \\
            0, & \textrm{$1 \leq [t]_T < n$} \nonumber \\
        \end{array} \right.
\\
    \psi_B (m,t) & = & 1
\end{eqnarray}
$T$ is the length of the schedule cycle and parameter $n$ equals to $T$.
Here we use $[t]_T$ notation to denote $t \mod T$, which is the slot index in a schedule cycle.

S-Nihao's bidirectional discovery is guaranteed since the interval of two consecutive beacons is fixed to the length of a slot,
and the single wakeup slot must overlap with at least one beacon wherever it is placed.
The behavior of S-Nihao is quite similar with broadcasting a packet in receiver-initiated MACs~\cite{sun2008rimac}\cite{dutta2010amac}, 
in which the sender will wake up for as long as the beacon period to ensure each neighbor will receive the broadcast.

S-Nihao's duty-cycle is:
\begin{equation} \label{snihaodc}
    DC = \frac{1 + \alpha (n-1)}{n} \approx \frac{1+ \alpha n}{n}
\end{equation}
For ease of calculation, we ignore one beacon slot that overlaps with wakeup slot, since $n$ is large enough in practice.

The discovery latency of S-Nihao is equal to $n$, \emph{i.e.}\ $L = n$, and the power-latency product is:
\begin{displaymath}
    \Lambda = DC \cdot L = \frac{1+\alpha n}{n} \cdot n = 1 + \alpha n
\end{displaymath}

Finally, we compare S-Nihao with the LL-Optimal discovery schedule:
\begin{equation} \label{compareoptimal}
    \lambda = \frac{\Lambda_{\phantom{1}}}{\Lambda_o} = \frac{1+ \alpha n}{\sqrt{n}} \geq \frac{2}{\sqrt{n}}
\end{equation}
when $\alpha=\frac{1}{n}$, Eqn. (\ref{compareoptimal}) gets the minimum value $\frac{2}{\sqrt{n}}$,
with $DC=\frac{2}{n}$ and $\Lambda=2$.

The result suggests that S-Nihao is even better than the LL-Optimal schedule when $n$ is sufficiently large.
Given a specific duty-cycle, S-Nihao will perform better with a lower latency bound.
For example, supposing the given duty-cycle is 5\%, LL-Optimal's latency is bounded by 400 slots, while S-Nihao's bound is 40 slots.
With larger $n$, S-Nihao is even more superior than LL-Optimal.
If the given duty-cycle is 1\%, LL-Optimal's worst-case latency bound is 10000 slots,
while 200 slots is enough in S-Nihao.

However, we have to note that $\alpha$ must be smaller than $DC$ in Eqn.~(\ref{snihaodc}), which is not easy to satisfy.
Suppose the desired duty-cycle is 1\% and the time for beacon transmission takes 1ms,
the time slot should be longer than 100ms to get $\alpha < 0.01$, which will increase the actual discovery latency.
We will improve S-Nihao to operate on any duty-cycle in the following text.

\subsection{Is TMLL Always True?}
We see S-Nihao greatly outperforms existing discovery protocols when only duty-cycle and latency are taken into consideration
(10 times better with DC=5\%, 50 times better with DC=1\%).
While the result is surprising, we wonder whether there is any additional cost paid for these improvements.
In the design of S-Nihao, we aggressively utilize the TMLL principle by sending beacons as more as possible, leaving only one wakeup slot in a schedule cycle.

The drawback of S-Nihao is that there may be too many beacons if a node is surrounded by plenty of neighbors, which involves extra interference on regular data communication and even the transmission of beacons.
It is reasonable that the fraction of beacons in a schedule cycle should be adjustable to meet different requirements.
As a result, we need a new metric that is never proposed by existing related works to quantify the number of transmitted beacons in a schedule cycle,
which is described in the next section.

\subsection{Channel Occupancy Rate}\label{channeloccupancy}
In this section, we introduce the channel occupancy rate metric to quantify the degree that a discovery protocol can occupy the channel.
Channel Occupancy Rate (COR) is the fraction of the time that the channel is occupied.
In a discovery schedule cycle, COR is defined as:
\begin{displaymath}
    COR = \frac{\alpha \cdot N_B}{T}
\end{displaymath}
where we use $N_B=\sum_{t=0}^{T-1} \psi_B(m,t)$ to represent the number of beacons in a schedule cycle.
Since each slot exists at most one beacon in Talk-Listen model, we use a simplified $\eta$ to represent the COR:
\begin{displaymath}
    \eta = \frac{COR}{\alpha} = \frac{N_B}{T}
\end{displaymath}

To evaluate the holistic performance of a neighbor discovery protocol,
we introduce another metric $A$ that is the product of duty-cycle, worst-case latency and COR.
For a given periodic discovery schedule, $A$ is defined as:
\begin{displaymath}
    A = DC \cdot L \cdot \eta 
\end{displaymath}

Now, we adopt the $A$ metric to analyze the performance of representative neighbor discovery protocols.
For the sake of clarity, we suppose the parameters are sufficiently large
so that the constant terms and floors/ceilings in original expressions can be omitted.

Quorum~\cite{tseng2003power} selects one row and one column for active slots.
If quorum size is $n^2$, $A$ is:
\begin{displaymath}
    A_Q = \frac{2}{n} \cdot n^2 \cdot \frac{2}{n} = 4
\end{displaymath}
U-Connect~\cite{kandhalu2010u} chooses a prime number $p$ for schedule with the row of active slots reduced by half. $A$ for U-Connect is:
\begin{displaymath}
    A_U = \frac{3}{2p} \cdot p^2 \cdot \frac{3}{2p} = \frac{9}{4} = 2.25
\end{displaymath}
SearchLight~\cite{bakht2012mobicom} schedules with a $t \times \frac{t}{2}$ matrix. Without striped probing, $A$ for SearchLight is:
\begin{displaymath}
    A_S = \frac{2}{t} \cdot \frac{t^2}{2} \cdot \frac{2}{t} = 2
\end{displaymath}
For SearchLight with 2-beacon approach and striped probing, discovery latency is reduced by half and $\eta$ is doubled due to the extra beacons.
Therefore, $A$ is:
\begin{displaymath}
    A_{S'} = \frac{2}{t} \cdot \frac{t^2}{4} \cdot \frac{4}{t} = 2
\end{displaymath}
For LL-Optimal, the schedule cycle is $n^2$ and the number of active slots is $n$. Its $A$ is:
\begin{displaymath}
    A_O = \frac{1}{n} \cdot n^2 \cdot \frac{1}{n} = 1
\end{displaymath}
For S-Nihao with cycle $n$, since each slot should transmit a beacon, so $\eta=1$ and $A$ is:
\begin{displaymath}
    A_{SN} = \frac{2}{n} \cdot n \cdot 1 = 2
\end{displaymath}

Table~\ref{ndcomparision} summarizes the features of neighbor discovery protocols in the symmetric case.
We observe that the $\eta$ for S-Nihao is a constant that cannot be adjusted according to user's requirements,
due to the fact that it has to send a beacon in each slot.
Although S-Nihao is better than LL-Optimal on power-latency product $\Lambda$,
it has a larger $A$ when considering the COR.
Those facts make us rethink the design of S-Nihao.
The core of the problem is to make $\eta$ adjustable.

\subsection{Generic Nihao}
\begin{figure}[t]
    \centering
    \subfigure[The schedule of G-Nihao in a line] {
        \includegraphics[scale=0.5]{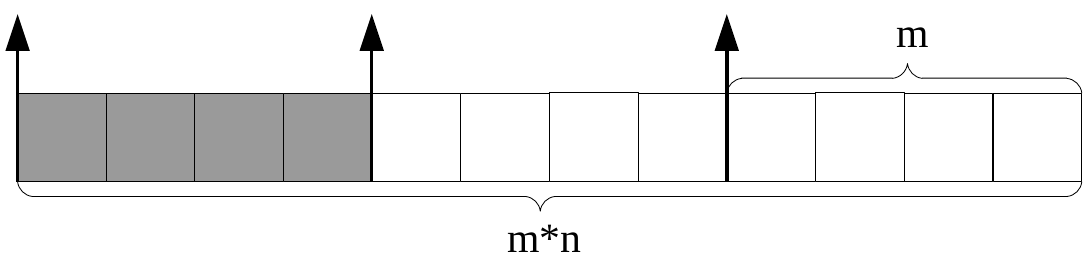}\label{nihao1}
    }
    \subfigure[Matrix representation of G-Nihao] {
        \includegraphics[scale=0.5]{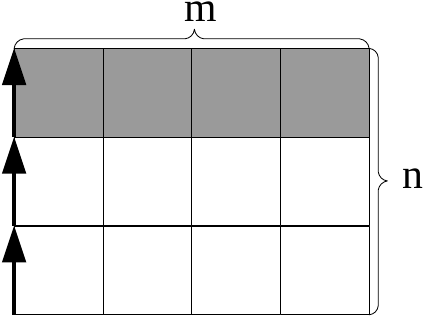}\label{nihao2}
    }
\caption{Two representations of G-Nihao.}
\end{figure}

This section presents the generic Nihao protocol (G-Nihao), the first one to consider COR in neighbor discovery,
whose $\eta$ can be flexibly adjusted to satisfy different requirements.
Instead of sending a beacon in each slot, G-Nihao is able to skip several slot as Fig.~\ref{nihao1} shows.
It is clearer to illustrate the schedule with the matrix representation in Fig.~\ref{nihao2}.

The schedules of G-Nihao can be defined as:
\begin{eqnarray}
    \psi_L (g,t) & = & \left\{ \begin{array}{ll}
            1, & \textrm{if $[t]_L < m$} \nonumber \\
            0, & \textrm{otherwise} \nonumber \\
        \end{array} \right.
\\
    \psi_B (g,t) & = & \left\{ \begin{array}{ll}
            1, & \textrm{$[t]_L=mi, i=0,1, \ldots, n-1$} \nonumber \\
            0, & \textrm{otherwise} \nonumber \\
        \end{array} \right.
\end{eqnarray}

The latency of G-Nihao is $L=mn$, and the duty-cycle is:
\begin{displaymath}
    DC=\frac{m + \alpha (n-1)}{mn} \approx \frac{m + \alpha n}{mn}
\end{displaymath}
We ignore one beacon as Eqn.~(\ref{snihaodc}) does for ease of calculation.
The power-latency product $\Lambda$ of G-Nihao is:
\begin{displaymath}
    \Lambda = \frac{m + \alpha n}{mn} \cdot mn = m + \alpha n
\end{displaymath}
Since the COR of G-Nihao is $\eta=\frac{n}{mn}=\frac{1}{m}$, the DC-L-COR product is:
\begin{displaymath}
    A = \frac{m + \alpha n}{mn} \cdot mn \cdot \frac{1}{m} = \frac{m + \alpha n}{m} = 1 +  \alpha \frac{n}{m}
\end{displaymath}

Since $\alpha \frac{n}{m}$ is adjustable by changing $m$ or $n$, G-Nihao can be better than all the existing discovery protocols except the LL-Optimal.
S-Nihao is a special case of G-Nihao with $m=1$.

\subsection{Balanced Nihao}
G-Nihao is a flexible protocol whose performance is determined by two parameters $m$ and $n$.
The user of G-Nihao should provide at least two precisely defined parameters among duty-cycle, latency and COR to calculate $m,n$.

\begin{figure}[t]
    \centering
    \includegraphics[scale=0.4]{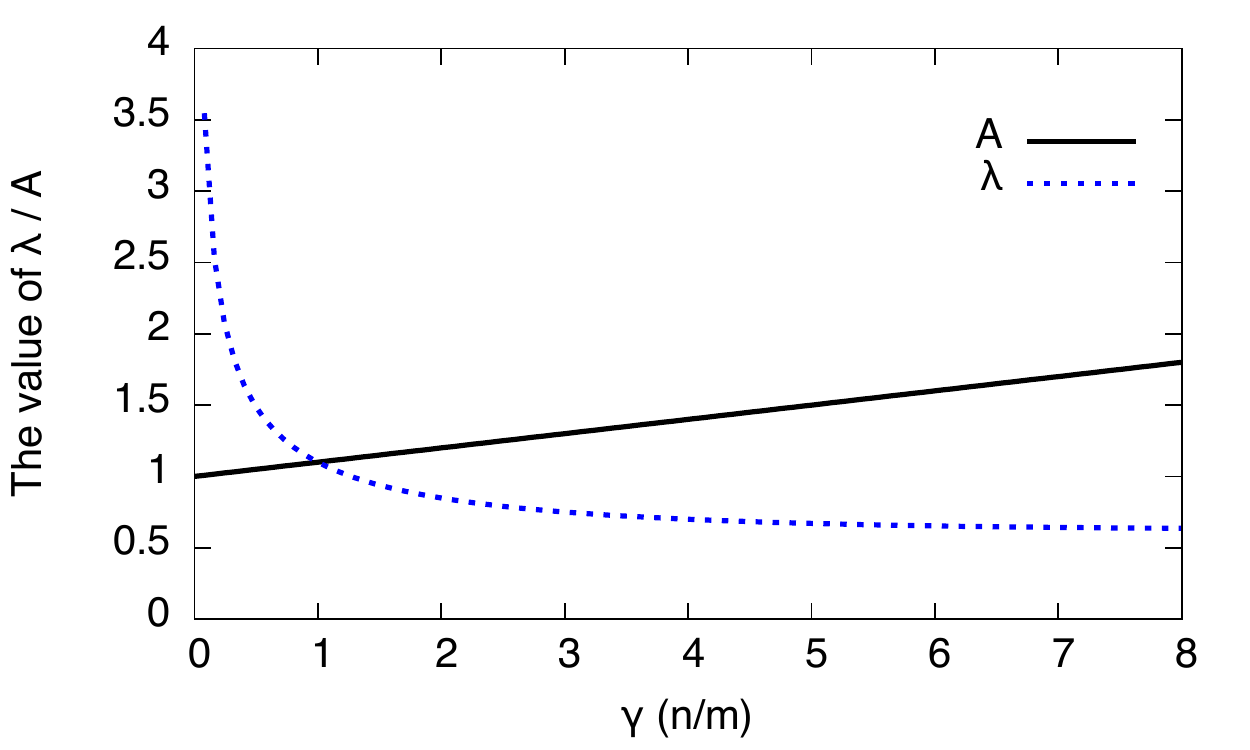}
    \caption{Find the balance between $\lambda$ and DC-L-COR product $A$. The balance is achieved when $\gamma=1$.}\label{apl}
\end{figure}

From a practical view, compared to duty-cycle and latency, COR is not easy to be defined clearly in applications.
We prefer to reduce the number of parameters in G-Nihao without sacrificing performance while keep COR reasonably low.
This problem is equivalent to finding the balance between $\lambda$ and $A$.
Figure~\ref{apl} shows the graph of the functions for $\lambda$ and $A$ with $\gamma=\frac{n}{m}$ and $\alpha=0.1$.
As $\gamma$ increases, $\lambda$ decrease quickly while $A$ increases slowly before they intersect.
The intersection point of two curves is the balance point for $\lambda$ and $A$.

We calculate the value of $\gamma$ for the intersection point:
\begin{eqnarray}
    \lambda = A & \Rightarrow & \frac{m+\alpha n}{\sqrt{mn}}=1+ \alpha \frac{n}{m} \nonumber \\
                & \Rightarrow & \frac{1}{\sqrt{\gamma}}+  \alpha \sqrt{\gamma} =  1 + \alpha \gamma \nonumber \\
                & \Rightarrow & \gamma = 1 \nonumber
\end{eqnarray}
It is interesting to find that the balance is achieved when $\gamma=1$, \emph{i.e.} $m=n$,
leading to a schedule that can be represented with a square matrix.
Therefore, we call $\gamma$ as the \emph{balance factor} that represents the degree of balance among duty-cycle, latency and COR.
More generally, we define $\gamma$ as:
\begin{displaymath}
    \gamma = \frac{N_B}{N_L}
\end{displaymath}
$N_L$ and $N_B$ are the number of wakeup and beacon slots respectively.
When $\gamma > 1$, the beacons are more likely to collide with each other, which will increase discovery latency.
In contrast, if $\gamma < 1$, discovery protocols will suffer from idle-listening with unnecessary active slots.

The final balanced Nihao's duty-cycle is $\frac{1+\alpha}{n}$ and worst-case latency is $n^2$, with the $\eta = \frac{1}{n}$.
Balanced Nihao (B-Nihao) is most suitable for practical applications in the symmetric case with the best-balanced performance.

\begin{table*}[t]
\centering
\caption{Comparison of neighbor discovery protocols}\label{ndcomparision}
\begin{tabular}{l|c|cccc|cccc|c}
\hline
       Protocol            & Parameter &        DC                   &         L         &  $\Lambda$              &           $\lambda$           &     $N_B$      &    $\eta$        &$\gamma$& $A$ & Asymm? \\
\hline                                                                                                                                                                                   
                                                                                                                                                                                         
Quorum                     &  $n$      & $\frac{2}{n}$               &  $n^2$            & $2\sqrt{L}-1$           & 2                             & $2n$           & $\frac{2}{n}$    & 1      & 4   & No \\ 
Disco                      &  $p_1,p_2$&$\frac{1}{p_1}+\frac{1}{p_2}$&  $p_1 p_2$        & $2\sqrt{L}$             & 2                             & $p_1+p_2$      & $\frac{1}{p_1}+\frac{1}{p_2}$    & 1      & 4   & Yes \\ 
U-Connect                  &  $p$      & $\frac{3}{2p}$              &  $p^2$            & $\frac{3\sqrt{L}+1}{2}$ & 1.5                           & $\frac{3p}{2}$ & $\frac{3}{2p}$   & 1      & 2.25& Yes \\
SearchLight                &  $t$      & $\frac{2}{t}$               &  $\frac{t^2}{2}$  & $\sqrt{2L}$             & $\sqrt{2}$                    & $t$            & $\frac{2}{t}$    & 1      & 2   & Yes \\
SearchLight (2B+stripe)    &  $t$      & $\frac{2}{t}$               &  $\frac{t^2}{4}$  & $\sqrt{L}$              & 1                             & $t$            & $\frac{4}{t}$    & 2      & 2   & Yes \\
BlindDate (4B+stripe)      &  $s$      & $\frac{3}{5s}$              &  $\frac{5s^2}{2}$ & $\sqrt{\frac{9}{10}L}$  & $\sqrt{\frac{9}{10}}$         & $6s$           & $\frac{12}{5s}$  & 4      & 3.6 & Yes \\
LL-Optimal (Combinatoric)  &  $n$      & $\frac{1}{n}$               &  $n^2$            & $\sqrt{L}$              & 1                             & $n$            & $\frac{1}{n}$    & 1      & 1   & No \\
Simplified Nihao           &  $n$      & $\frac{2}{n}$               &  $n$              & 2                       & $\frac{2}{\sqrt{n}}$          & $n$            &        1         & $n$    & 2   & Yes \\
Generic Nihao              &  $m,n$    & $\frac{m + \alpha n}{mn}$   &  $mn$             & $m+\alpha n $           & $\frac{m+\alpha n}{\sqrt{mn}}$& $n$            & $\frac{1}{m}$    & $\frac{n}{m}$ & $1+ \alpha \frac{n}{m}$ & Yes \\
Balanced Nihao             &  $n$      & $\frac{1+\alpha}{n}$        &  $n^2$            & $n(1+\alpha)$           & $1+\alpha$                    & $n$            & $\frac{1}{n}$    & 1      & $1+\alpha$ & No \\

\hline
\end{tabular}
\end{table*}

\subsection{Theoretical Bound for $A$}
As table~\ref{ndcomparision} shows, LL-Optimal is the best on the $A$ metric among the listed protocols.
We wonder whether LL-Optimal is still optimal in the Talk-Listen model.
If not, what is the bound of $A$?
We dive into these questions in this section.

\begin{figure}[t]
    \centering
    \includegraphics[scale=0.5]{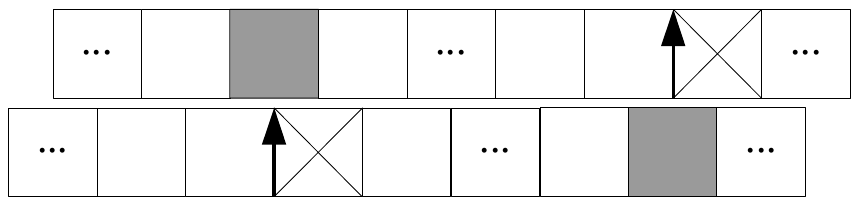}
    \caption{Convert Listen-Listen model to Talk-Listen model.
    With a discovery schedule ensuring two overlaps in Listen-Listen model, 
    bidirectional discovery in Talk-Listen model can be converted by substituting a wakeup slot to a beacon.
    The time slot marked with `X' is a wakeup slot in original schedule, which is converted to a beacon slot.
}\label{bidicoverytlmodel}
\end{figure}

To guarantee a bidirectional discovery in Talk-Listen model, there must be two overlaps that satisfy Eqn.~(\ref{bidiscoverydef}).
We are able to convert Listen-Listen model to Talk-Listen model by substituting wakeup slots with beacon slots as Fig.~\ref{bidicoverytlmodel} illustrates.
Given a cycle $T$ and the number of overlaps $m$, 
Zheng \emph{et al.}~\cite{zheng2003asynchronous} prove that the number of active slots in Listen-Listen model is $k \geq \sqrt{m \cdot T}$.
To ensure two overlaps, \emph{i.e.}\ $m=2$, the number of active slots satisfies:
\begin{equation}\label{eqnk}
 k \geq \sqrt{2T}
\end{equation}

Assuming we build an LL-Optimal schedule $\psi_o$ with at least 2 overlaps, and convert it to a $(\psi_L, \psi_B)$ schedule by substituting wakeup slots with beacon slots.
The number of active slots $k$ satisfies:
\begin{equation}\label{eqnk2}
    k = N_L + N_B
\end{equation}
$N_L$ and $N_B$ are the number of wakeup and beacon slots respectively.
Now, we can calculate the bound for $A$ with Eqn. (\ref{eqnk}) and Eqn. (\ref{eqnk2}):
\begin{eqnarray}\label{boundofa}
    A & =    & DC \cdot (L \cdot \eta) = \frac{N_L}{T} \cdot N_B    \nonumber \\
      & =    & \frac{N_B (k - N_B)}{T}                              \nonumber \\
      & \geq & \frac{N_B (\sqrt{2T} - N_B)}{T}  =  -(\frac{N_B}{\sqrt{T}})^2 + \sqrt{2}\frac{N_B}{\sqrt{T}}
\end{eqnarray}
Eqn.~\ref{boundofa} is a quadratic function with maximum value $\frac{1}{2}$ when $\frac{N_B}{\sqrt{T}}=\frac{\sqrt{2}}{2}$.
Thus, we get the lower bound for $A$ is:
\begin{displaymath}
A \geq \frac{1}{2}
\end{displaymath}
We notice that $N_B=N_L$, \emph{i.e.} $\gamma=1$ when $A$ gets the minimum value,
which implies a balance between the beacons and wakeup slots in the optimal schedule.

It is important to note that we haven't constructed the optimal schedule but just assume its existence.
It remains an open research problem as it is unknown whether such optimal schedule actually exists.
We will leave it as the future work.

\subsection{Asymmetry}
\begin{figure}[t]
    \centering

    \subfigure[$n_1=3$] {
        \includegraphics[scale=0.55]{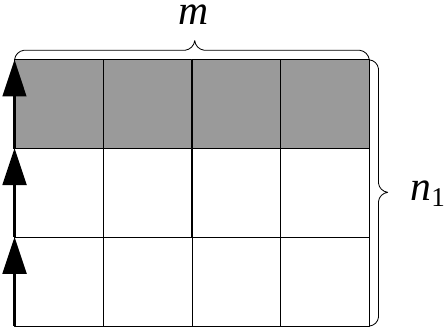}
    }
    \subfigure[$n_2=6$] {
        \includegraphics[scale=0.55]{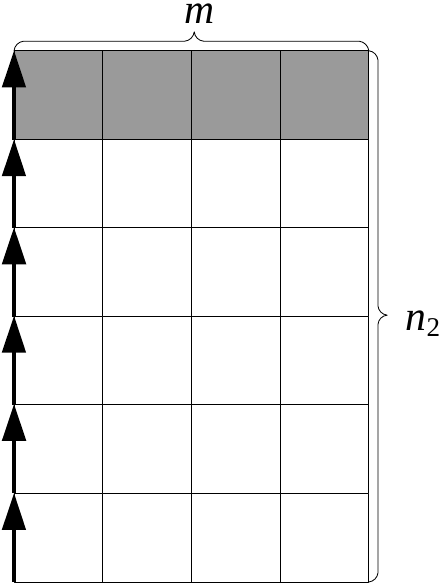}
    }
    \caption{G-Nihao for the asymmetric case.
        The value of $m$ should be the same to ensure discovery.
        $n$ is tunable for different duty-cycles.
        Here is an example for 33.3\% and 16.7\% duty-cycles with $m=4$, $n_1=3$ and $n_2=6$.}\label{asymmetricnihao}
\end{figure}

Since heterogeneous devices in sensor networks are likely to have diverse energy budgets,
it is reasonable that they operate with different duty-cycles.
G-Nihao supports the asymmetric case by simply adjusting the $n$ value according to the required duty-cycle.
As Fig.~\ref{asymmetricnihao} shows, G-Nihao guarantees discovery in the asymmetric case with different $n_1$ and $n_2$, while $m$ remains constant.
The duty-cycle is $\frac{1}{n_1}$, $\frac{1}{n_2}$ respectively,
and the worst-case latency is $m \cdot \max\{n_1,n_2\}$.

G-Nihao has a good \emph{duty-cycle granularity} in the asymmetric case.
Duty-cycle granularity is a notion proposed by Chen \emph{et al.}~\cite{chen2015infocom} that means how closely a discovery protocol can match the required duty-cycle.
In co-prime based protocols such as Disco and U-Connect, the duty-cycle is decided by a prime number $p$, which determines the duty-cycle to $\frac{1}{p}$.
Since the number of prime numbers is quite limited, it performs poorly on duty-cycle granularity.
SearchLight~\cite{bakht2012mobicom} restricts the duty-cycle to a power-multiple of the smallest duty-cycle in the asymmetric case, leading to even more limited choices.
Hedis and Todis~\cite{chen2015infocom} support duty-cycle in the form of $\frac{2}{n}$ and $\frac{3}{n}$ respectively, which have a better granularity.
G-Nihao is comparable to Hedis and Todis with a fine-grained duty-cycle in the form of $\frac{1}{n}$.

Given different duty-cycles, G-Nihao has multiple choices of parameters.
Since the value $m$ has to be the same in the network, balance cannot be achieved for different duty-cycles.
The question is what $m$ should be to balance the performance better in a global view.
We define the \emph{global balance factor} $\Gamma$ to answer this question:
\begin{displaymath}
    \Gamma = \prod_{i=1}^{d} \gamma_i
\end{displaymath}
$d$ is the number of different duty-cycles and $\gamma_i$ is the balance factor for the schedule of the $i$th duty-cycle.
Global balance is achieved when $\Gamma=1$, which means the number of beacons and wakeup slots are balanced.

For example, the desired duty-cycles are 1\% and 5\%.
Suppose $\alpha$ is sufficiently small, then $n_1=100$, $n_2=20$ for 1\% and 5\% duty-cycles respectively.
The global balance factor is:
\begin{eqnarray}
    \Gamma & = & \frac{n_1}{m} \cdot \frac{n_2}{m} = 1 \nonumber \\
           & \Rightarrow & m^2 = n_1 n_2 \nonumber \\
           & \Rightarrow & m = \sqrt{n_1 n_2} = 44.72 \nonumber
\end{eqnarray}
So we choose $m=45$ to achieve the best balance in such an asymmetric case.

\section{Evaluation}\label{evaluation}
In this section, we perform experiments on real-world testbeds to evaluate Nihao's performance.

\subsection{Implementation and Experiment Setup}
We have implemented Nihao and other reference protocols on TinyOS 2.1.2.
Beacons are implemented as small AM broadcast messages with zero payloads.
The length of the underlying physical message is 17 bytes (4B preamble + 1B SFD + 1B PHR + 8B MAC header + 1B TinyOS AM type + 2B CRC).
It takes 0.54ms for an IEEE 802.15.4-compatible radio to transmit the beacon.

Our implementation directly controls the radio on and off by invoking 
the \texttt{start()} and \texttt{stop()} commands in the \texttt{SplitControl} interface of \texttt{ActiveMessageC}.
We have observed that when the node is busy receiving messages,
the function call to \texttt{stop()} will return \texttt{EBUSY}, leaving the radio keeping on listening.
We resolve this problem by re-posting the stop task when a failure occurs, which will finally turn radio off.

We use 40 nodes in our evaluation.
The node consists of an ATMega128RFA1 MCU and a Serial-to-USB chip.
The MCU has 16KB RAM and operates on 16MHz, which integrates an IEEE 802.15.4-compatible radio.
We collect discovery events from the Serial-to-USB port without occupying the wireless channel.
Each discovery event is timestamped for calculating the cumulative distribution functions (CDFs) of discovery latency.
Although Serial-to-USB transmission delay is not considered in the timestamp,
it will not affect the accuracy of the result,
because nodes are directly connected to PC and the delay will be eliminated when the relative time is actually used for calculating the latency.

Nodes are placed closely to avoid interference, and any pair of nodes can communicate with each other (bidirectional link).
Time slot is set to 10ms as Disco does. As a result, $\alpha$ is $0.54/10=0.054$ for Nihao.
To avoid synchronization of wakeup schedules, we add some random delay before nodes start discovering neighbors, and the seed of random number generator is the node's ID.

\subsection{Redundant Beacon for Bidirectional Discovery}\label{redundantbeaconexpsection}
\begin{figure}[t]
    \centering
    \includegraphics[scale=0.5]{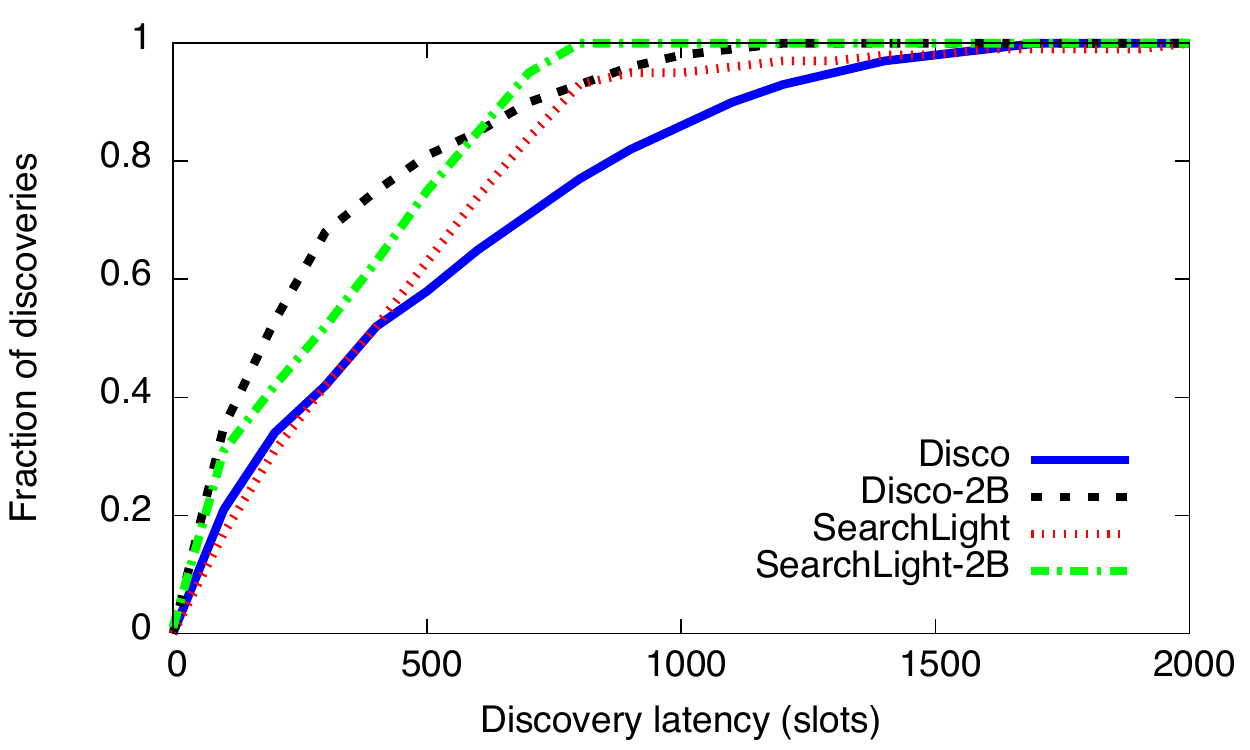}
    \caption{2-beacons approach is redundant. The two versions of Disco and SearchLight are evaluated with 5\% duty-cycle.
             Both of them reach 100\% within the worst-case bound.
             2-beacon versions are slightly faster.}\label{2beacondup}
\end{figure}

We have claimed in section~\ref{redundantbeaconsection} that one beacon is enough for bidirectional discovery.
Now we validate it with experiments. 
We test Disco with 1-beacon and 2-beacon in active slot respectively, so does SearchLight.
The 2-beacon version is named as Disco-2B and SearchLight-2B.
As Fig.~\ref{2beacondup} shows, both versions reach 100\% before the worst-case latency, which is consistent with our proof.
However, 2-beacon versions have slightly better discovery rate, which have inspired us to add more beacons to reduce latency.

\subsection{Symmetric Discovery}
\begin{figure*}[t]
    \centering
    \begin{minipage}[t]{0.3\textwidth}
    \centering
    \includegraphics[scale=0.45]{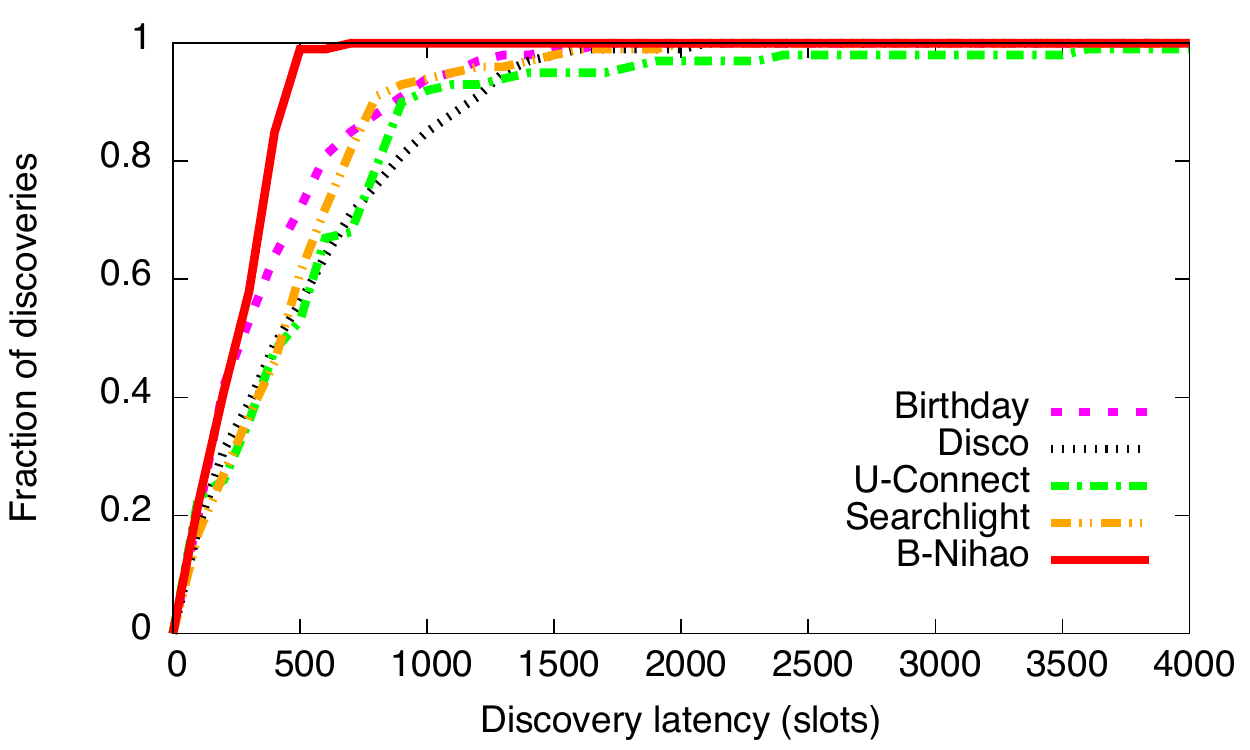}
    \caption{CDFs of discovery latency with duty-cycle 5\%.}\label{dc5}
    \end{minipage}
    \hspace{1em}
    \begin{minipage}[t]{0.3\textwidth}
    \centering
    \includegraphics[scale=0.45]{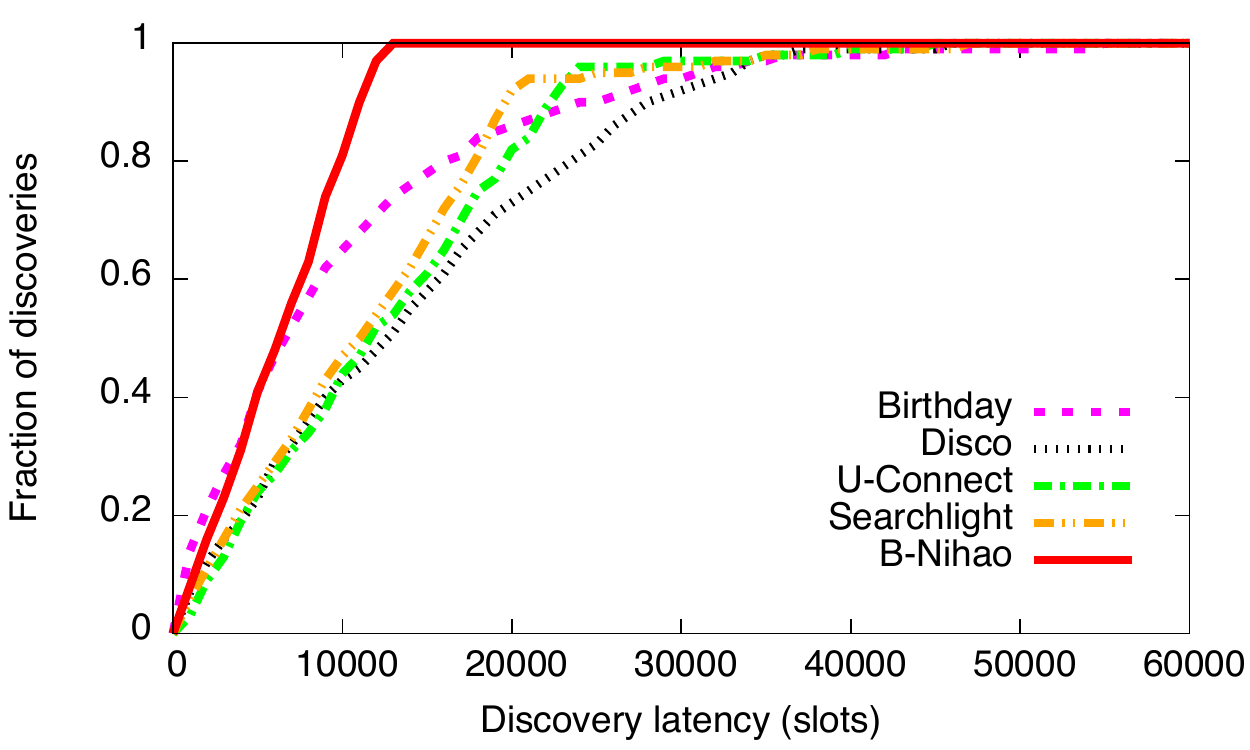}
    \caption{CDFs of discovery latency with duty-cycle 1\%.}\label{dc1}
    \end{minipage}
    \hspace{1em}
    \begin{minipage}[t]{0.3\textwidth}
    \centering
    \includegraphics[scale=0.45]{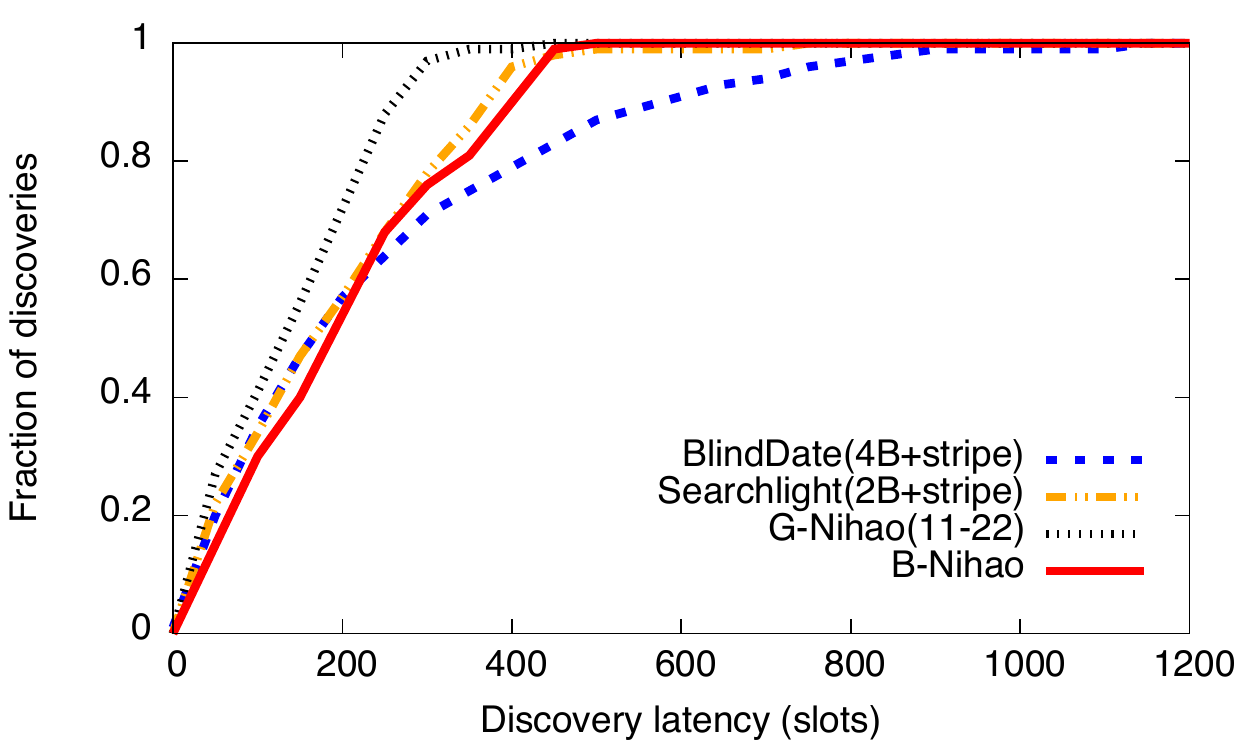}
    \caption{Compare Nihao with striped SearchLight and BlindDate at the duty-cycle of 5\%.}\label{slbd}
    \end{minipage}
\end{figure*}

We evaluate the performance of B-Nihao and compare it with other protocols in the symmetric case.
The reference protocols are the 1-beacon version for fairness, which means they send only one beacon in the active slot.
The balance factor $\gamma=1$ for all protocols including B-Nihao. 
The result is depicted in Fig.~\ref{dc5} and Fig.~\ref{dc1} for 5\% and 1\% duty-cycle respectively.
B-Nihao is significantly faster than the others in both cases with lowest latency bound.
It's worth noting that other protocols don't reach 100\% in both cases due to beacon collisions.

Now we compare Nihao with striped SearchLight (\emph{i.e.} SearchLight with striped probing) and BlindDate, which are reported to have the best performance in the symmetric case so far.
Since striped SearchLight has to send two beacons in each active slot, the balance factor for striped SearchLight is $\gamma=2$.
Similarly, $\gamma$ is 4 for BlindDate which sends two more beacons.
To fairly compare with striped SearchLight, we choose G-Nihao with $\gamma=2$ ($m=11, n=22$) when duty-cycle is 5\%.
We also test B-Nihao with the same duty-cycle ($n=21$) to get a better understanding of these protocols.

The result in Fig.~\ref{slbd} suggests that G-Nihao is faster than striped SearchLight and BlindDate since it has the lowest worst-case latency (G-Nihao:242, SearchLight:400, BlindDate:360, B-Nihao:441).
We also note that the protocols with $\gamma > 1$ can't reach 100\% discovery within worst-case latency, which is caused by beacon collision.
BlindDate suffers most from beacon collision since it has the largest $\gamma=4$.
In contrast, B-Nihao is the first to reach 100\% discovery with least collision, although it is not as fast as the other protocols.

\subsection{Asymmetric Discovery}
\begin{figure}[t]
    \centering
    \includegraphics[scale=0.5]{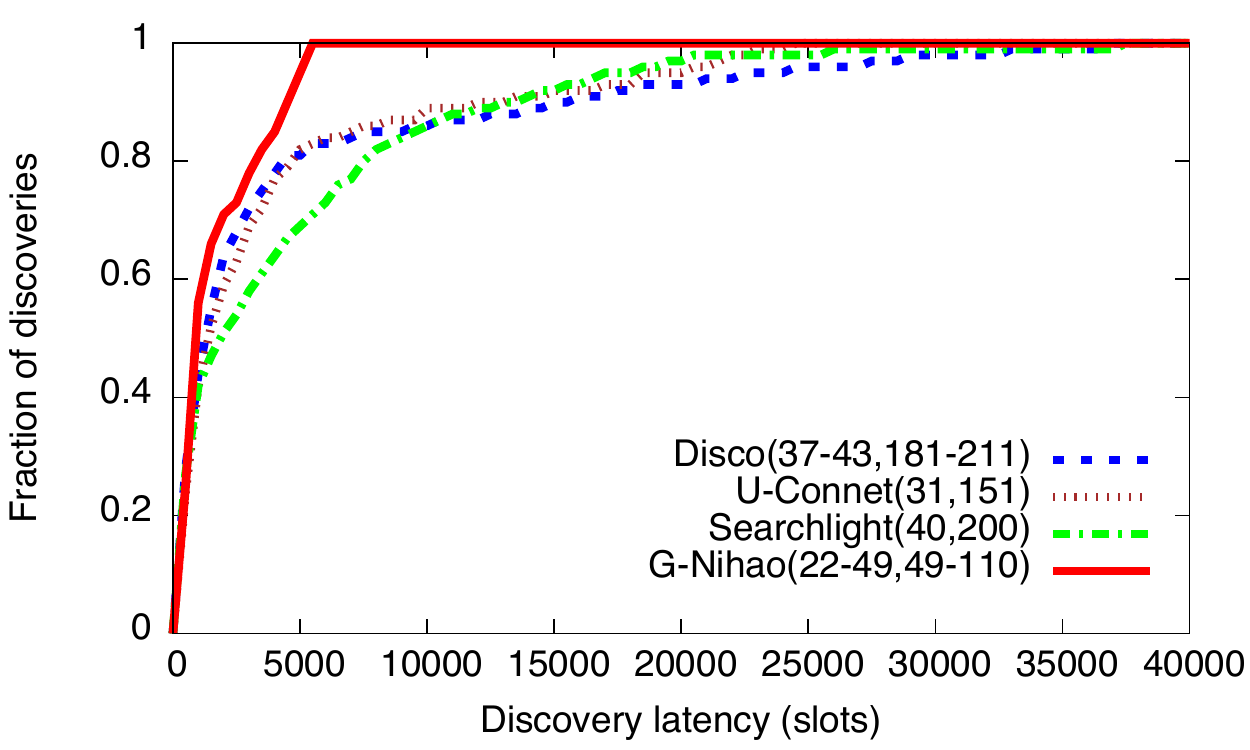}
    \caption{CDFs of discovery latency in the asymmetric case. Two group of nodes operate at 1\% and 5\% duty-cycles respectively. Parameters for corresponding protocols are exhibited in parentheses.}\label{asymexp}
\end{figure}

Since B-Nihao only works in symmetric case, we evaluate the performance of G-Nihao with proper parameters and compare it with other protocols under asymmetric duty-cycles.
We divide 40 nodes into 2 equal groups. Nodes in one group operate with 1\% duty-cycle, and the others operate with 5\% duty-cycle.
Four protocols with appropriate parameters are tested in this scenario. 
The parameters for G-Nihao are $n_1=110$ for 1\% duty-cycle and $n_2=22$ for 5\% duty-cycle, while $m$ is $\sqrt{n_1 n_2} \approx 49$ to make global balance factor $\Gamma=1$.
Fig.~\ref{asymexp} shows the result, when the fraction of discoveries exceeds 50\%,
the discovery rate slows down since nodes with 5\% duty-cycle have discovered each other with bounded latency.
G-Nihao is the winner among the four protocols reaching 100\% discovery with about 5500 slots, which is much faster than the other reference protocols.
The reason is G-Nihao has a much smaller worst-case bound (12100) than the others (Disco:38191,U-Connect:22801, SearchLight:20000).

\section{Related Works}\label{relatedworks}
A wide range of neighbor discovery protocols have been proposed for wireless sensor networks in recent years.
Since nodes are not necessarily synchronized with each other and precise global time synchronization is expensive,
asynchronous neighbor discovery is preferred in practical applications.

Existing asynchronous neighbor discovery protocols can be roughly divided into two categories: probabilistic and deterministic~\cite{sun2014ndsurvey}.
Birthday~\cite{mcglynn2001birthday} is a probabilistic neighbor discovery protocol using random independent transmissions to find neighbors.
Each time slot randomly chooses one state among listen, transmit and idle with probabilities which is determined by the required duty-cycle.
Although Birthday protocol has an excellent performance on average discovery latency thanks to its probabilistic nature, 
it exhibits a long tail for discovering the last few neighbors, and the worst-case latency is unbounded.

To make worst-case latency bounded, various deterministic neighbor discovery protocols are developed.
Quorum~\cite{tseng2003power} protocol utilizes the concept of quorum widely used in distributed systems to guarantee discovery between a pair of nodes.
It rearranges the continuous $n^2$ slots to a schedule matrix and chooses one row and one column as the active slots,
which will definitely overlap regardless of the offset between two schedules.
The problem of Quorum is it has two overlap slots while the latter one is redundant for discovery.

U-Connect~\cite{kandhalu2010u} presents a better wakeup schedule than Quorum with fewer active slots.
By providing a Listen-Listen model of the asynchronous neighbor discovery problem,
U-Connect proves that it is 1.5-approximation to the optimal using the power-latency product metric.
We have used similar notations with U-Connect to formalize performance metrics and
extend the Listen-Listen model to better analyze the wakeup schedules that beacons are not inside the wakeup slots.

SearchLight~\cite{bakht2012mobicom} designs a more effective wakeup schedule that achieves $\sqrt{2}$-approximation to the optimal.
The author also claims that striped probing is helpful to reduce worst-case latency, which has inspired us to transmit more beacons to reduce probes.
BlindDate~\cite{wang2015blinddate} achieves an even lower latency bound by adding two extra beacons, giving us the intuition that beacons are not necessarily placed in wakeup slots.

Hello~\cite{sun2014hello} presents a generic framework for quorum-based discovery protocols such as Quorum, U-Connect, Disco and SearchLight in symmetric cases.
It is flexible in adjusting parameters to better meet various demands. Our G-Nihao exhibits the same flexibility as Hello,
while has a better performance.

Zheng~\cite{zheng2003asynchronous} puts a theoretical bound for symmetric discovery protocols in Listen-Listen model by combinatoric design.
We have applied the theorem provided by the author to prove the lower bound of DC-L-COR product in our extended Talk-Listen model.

Asymmetric neighbor discovery is proposed for cases that duty-cycle requirements are different.
Lai \emph{et al.} extend Quorum to support the asymmetric case of two independent duty-cycles.
Disco~\cite{dutta2008practical} utilizes the Chinese Remainder Theorem to guarantee discovery between two nodes with different duty-cycles.
U-Connect and Hello employ similar coprime approach for asymmetric cases.
SearchLight and BlindDate adopt another power-multiple approach with a poor duty-cycle granularity.
Hedis and Todis~\cite{chen2015infocom} present wakeup schedules with better duty-cycle granularities for asymmetric cases.
Our design supports asymmetric cases with the same duty-cycle granularity as Todis and Hedis, but has a better performance on latency.

Collaborative neighbor discovery~\cite{zhang2012acc}\cite{cohen2011continuous}\cite{chen2012secon}\cite{purohit2011wiflock}
is proposed to accelerate the discovery process by exchanging neighbor information, exploiting the fact that neighbor relationship is transitive.
Although we only focus on the basic independent neighbor discovery problem, the approach proposed in this paper is also applicable to collaborative methods,
such as Acc~\cite{zhang2012acc}, which is an on-demand accelerating middleware for existing neighbor discovery protocols.

\section{Conclusion}\label{conclusion}
We have presented Nihao, a family of energy-efficient asynchronous neighbor discovery protocols.
G-Nihao is a flexible protocol for both symmetric and asymmetric cases with two tunable parameters.
B-Nihao is a symmetric protocol with the best-balanced performance while involes only one parameter.
The analytical and real-world experiment results show that Nihao is significantly better than the state of the art protocols.

We come up with two basic ideas to design Nihao.
One is beacons are not necessarily placed in active slots to reduce idle-listening.
The other is we have considered the DC-L-COR product and balance factor to trade off among duty-cycle, latency and channel occupancy rate.
We believe these ideas will be helpful in designing future neighbor discovery protocols.

\section*{Acknowledgment}
This work was supported by National Key Technologies R\&D Program of China Grant No. 2014BAH14F01,
National S\&T Major Project of China Grant No. 2012ZX03005007, National NSF of China Grant No. 61402372 and Fundamental Research Funds for the Central Universities Grant No. 3102014JSJ0003.

\bibliographystyle{IEEEtran}
\bibliography{ref}
\end{document}